\documentclass{aastex63}

\usepackage{amsmath}
\usepackage{url}
\usepackage{soul}
\usepackage{float}
\usepackage{bm}

\submitjournal{The Astrophysical Journal}
\accepted{November 16, 2020}
\shorttitle{High Mach number Shocks with MMS}
\shortauthors{Madanian et al.}
\graphicspath{{./}{figures/}}
\begin{document}

\title{The Dynamics of a High Mach Number Quasi-Perpendicular Shock: MMS Observations}


\correspondingauthor{Hadi Madanian}
\email{hmadanian@swri.edu}

\author[0000-0002-2234-5312]{H. Madanian}
\affiliation{Southwest Research Institute, 6220 Culebra Rd, San Antonio, TX 78238, USA}

\author{M.I. Desai}
\affiliation{Southwest Research Institute, 6220 Culebra Rd, San Antonio, TX 78238, USA}
\affiliation{University of Texas at San Antonio, San Antonio, TX 78249, USA}

\author{S.J. Schwartz}
\affiliation{Laboratory for Atmospheric and Space Physics, University of Colorado, Boulder, CO 80303, USA}

\author{L.B. Wilson III}
\affiliation{NASA Goddard Space Flight Center, Greenbelt, MD 20771, USA}

\author{S.A. Fuselier}
\affiliation{Southwest Research Institute, 6220 Culebra Rd, San Antonio, TX 78238, USA}
\affiliation{University of Texas at San Antonio, San Antonio, TX 78249, USA}

\author{J.L. Burch}
\affiliation{Southwest Research Institute, 6220 Culebra Rd, San Antonio, TX 78238, USA}

\author{O. Le Contel}
\affiliation{Laboratoire de Physique des Plasmas, CNRS, Ecole Polytechnique, Sorbonne Université, Université Paris-Saclay, Observatoire de Paris, Paris, France}

\author{D.L. Turner}
\affiliation{Johns Hopkins University Applied Physics Laboratory, Laurel, MD 20723, USA}

\author{K. Ogasawara}
\affiliation{Southwest Research Institute, 6220 Culebra Rd, San Antonio, TX 78238, USA}

\author{A.L. Brosius}
\affiliation{NASA Goddard Space Flight Center, Greenbelt, MD 20771, USA}
\affiliation{Pennsylvania State University, University Park, PA 16802, USA}

\author{C.T. Russell}
\affiliation{University of California, Los Angeles, CA 90095, USA}

\author{R.E. Ergun}
\affiliation{Laboratory for Atmospheric and Space Physics, University of Colorado, Boulder, CO 80303, USA}

\author{N. Ahmadi}
\affiliation{Laboratory for Atmospheric and Space Physics, University of Colorado, Boulder, CO 80303, USA}

\author{D.J. Gershman}
\affiliation{NASA Goddard Space Flight Center, Greenbelt, MD 20771, USA} 

\author{P.-A. Lindqvist}
\affiliation{KTH Royal Institute of Technology, Stockholm 10044, Sweden}

\begin{abstract}
Shock parameters at Earth's bow shock in rare instances can approach the Mach numbers predicted at supernova remnants. We present our analysis of a high Alfv\'en Mach number ($M_A= 27$) shock utilizing multipoint measurements from the Magnetospheric Multiscale (MMS) spacecraft during a crossing of Earth's quasi-perpendicular bow shock. We find that the shock dynamics are mostly driven by reflected ions, perturbations that they generate, and nonlinear amplification of the perturbations. Our analyses show that reflected ions create modest magnetic enhancements upstream of the shock which evolve in a nonlinear manner as they traverse the shock foot. They can transform into proto-shocks that propagate at small angles to the magnetic field and towards the bow shock. The nonstationary bow shock shows signatures of both reformation and surface ripples. Our observations indicate that although shock reformation occurs, the main shock layer never disappears. These observations are at high plasma $\beta$, a parameter regime which has not been well explored by numerical models.
\end{abstract}


\section{\label{sec:Intro}Introduction}

The physics of collisionless shocks have been extensively investigated over the past several decades through theoretical models, numerical simulations, in-situ observations, and laboratory experiments \cite[\ and references therein]{Bykov2019ShocksGalaxies, Parks2017ShocksPlasmas, Balogh2013PhysicsWaves, Schaeffer2017High-MachShocks, burgess_collisionless_2015,Bell2014ParticleRemnants, Treumann2009, lembege_selected_2004, Gedalin1997IonFront}. Collisionless shocks in space plasmas are characterized by several key parameters including Mach number, which specifies the flow speed relative to the phase speed of a characteristic wave mode in the background plasma. Plasma $\beta$, or the ratio of the thermal to magnetic pressures, of the incident flow is also an important parameter that can affect the growth rate of various plasma instabilities. A third important parameter is the angle ($\theta_{Bn}$) that the upstream magnetic field makes with the shock normal. Shocks exhibit significantly different characteristics based on $\theta_{Bn}$. At quasi-perpendicular shocks ($\theta_{Bn}>45^\circ$), the guiding center of reflected charged particles is driven towards the shock by the solar wind so their trajectories are constrained to about one gyroradius distance upstream. In contrast, at quasi-parallel shocks ($\theta_{Bn}<45^\circ$), particles can gyrate and stream along the magnetic field line farther upstream. These differences in trajectories cause different particle distributions, plasma instabilities, and shock structures \citep{Burgess2006Quasi-parallelProcesses, bale_quasi-perpendicular_2005}. The angle $\theta_{Bn}=45^\circ$ is conveniently chosen to separate these regimes, though in some relativistic shocks the quasi-perpendicular regime begins at much smaller angles ($\theta_{Bn} > 34^\circ$) \citep{Bykov2011FundamentalsShocks}. For the rest of the paper we shall focus on the quasi-perpendicular regime. 

In supercritical shocks, wave-particle coupling within the transition layer is insufficient to dissipate the incident kinetic energy and to sustain a stable shock layer. The energy balance at these shocks is attained by radiating dispersive whistler waves \citep{Fairfield1974WhistlerShocks, tidman_emission_1968,wilson_observations_2012, wilson_revisiting_2017} and reflecting a portion of the solar wind ions upstream \citep{Paschmann1982, Schwartz1983IonsPopulations}. At higher Mach numbers, unbalanced nonlinear growth of the shock layer also leads to a nonstationary shock front. We will consider two forms of nonstationarity: 1) shock front reformation, and 2) surface ripples. Two common models that describe the reformation process are as follows: Shock reformation can be caused by nonlinear steepening of whistler waves at the shock ramp via the so-called gradient catastrophe process \citep{krasnoselskikh_nonstationarity_2002, lobzin_nonstationarity_2007, galeev_physical_1988, dimmock_direct_2019}. This is a 1D approximated model which does not include dissipation effects due to reflected ions. Reformation can also be mediated by the dynamics of reflected ions in the foot region. Numerical simulations have shown that accumulation of reflected ions at the upstream edge of the foot gives rise to a new shock front which replaces the old shock upon formation \citep{biskamp_numerical_1972, hada_shock_2003,lembege_nonstationarity_1992, lembege_nonstationarity_2009, scholer_quasi-perpendicular_2003, hellinger_reformation_2002}. These models predict shock reformation at low $\beta$ ($\leq1$) plasmas only. Due to numerical constraints, and unrealistic ion to electron mass ratio and plasma to cyclotron frequency ratio, dispersive effects in these models are highly overestimated. The reformation process has a period on the order of solar wind proton gyroperiod. 

When the upstream Mach number is higher than the nonlinear whistler critical Mach number, shock-generated whistler waves cannot propagate in the foot, or phase stand in the solar wind as a precursor or as an isolated soliton \citep{krasnoselskikh_dynamic_2013, galeev_physical_1988}. In this case, the dynamics of the shock layer are driven by reflected ions and instabilities they generate. Reflected ions cause significant temperature anisotropy and can initiate instabilities that excite dispersive waves acting to relax the background anisotropy \citep{Winske1988MagneticShocks}. Surface ripples have been associated with large amplitude low frequency waves within the shock ramp and overshoot excited by Alv\'en Ion Cyclotron (AIC) instability \citep{lowe_properties_2003, Yang2018ImpactShocks, Burgess2016MicrostructureShocks, Moullard2006RipplesShock, johlander_rippled_2016,gingell_mms_2017, johlander_shock_2018, hanson_crossshock_2019}. The ripple waves propagate along the shock surface at speeds comparable to the local Alfv\'en speed. Rippling alters the orientation of the magnetic field across the shock most noticeably observed as oscillations in the normal component. 

Changes in the upstream solar wind conditions can also lead to displacement of the entire bow shock layer. This form of nonstationarity typically occurs on timescales larger than the characteristic timescales of the charged particle dynamics.

Our goal in this study is to provide observational details of the shock layer dynamics at high Alfv\'en Mach numbers ($M_A > 25$). This is a parameter regime close and somewhat comparable to weak shock waves at supernovae remnants (SNRs) and other astrophysical structures \citep{masters_comparison_2013, Petrukovich2019Low-frequencyShock, Ghavamian2013Electron-ionConnection}. In-situ observational studies at Earth's bow shock can provide boundary conditions and establish constraints on models that describe SNR shocks, which are considered the primary source of cosmic rays. Higher magnetic amplification across the shock can lead to higher particle acceleration rates \citep{Caprioli2014SIMULATIONSAMPLIFICATION, Caprioli2014SIMULATIONSEFFICIENCY, Bykov2011CosmicInstability,Bell2004TurbulentRays,Bell2001CosmicField,Lucek2000Non-linearStreaming}. It has been shown that the magnetic amplification rate increases with upstream Mach number \citep{Russell1982OvershootsShocks}, and plasma $\beta$ \citep{Winterhalter1988ObservationsConditions}. Nonstationary shocks that undergo reformation show even higher amplification rates \citep{sulaiman_quasiperpendicular_2015}, highlighting the importance of the shock structure dynamics in particle acceleration. Current models of nonstationarity however, are inconsistent and insufficient in describing the shock dynamics at high Mach numbers \citep{sundberg_dynamics_2017}. We take a somewhat different approach to shock reformation than the classical view of this process. That is, similar upstream disturbances or accumulation of reflected ions is seen, but we still see ion signatures due to reflection at the pre-existing shock. The role of ions in driving the shock layer dynamics and generating upstream instabilities is emphasized.

\section{Data and Methods}
\label{sec:meth}
\subsection{Data}
We use multipoint measurements of Magnetospheric Multiscale (MMS) spacecraft \citep{Burch2016MagnetosphericObjectives}, during a burst mode operation interval when data from all instruments are at their highest resolution. Magnetic field data are from the fluxgate magnetometer (FGM) instrument with a measuring cadence of 128 samples per second \citep{Russell2016TheMagnetometers}. The ion and electron data are from the Fast Plasma Investigation (FPI) instrument which provides ion and electron measurements of the full sky field-of-view (FOV) in the 10 eV/q to 30 keV/q energy-per-charge range \citep{pollock_fast_2016}. In the burst mode, FPI samples electron and ion populations every 30 and 150 milliseconds, respectively. Ion moments are calculated for solar wind and reflected ions separately. Upstream of the bow shock, reflected and scattered ions are detected in different angles than the unshocked solar wind beam. This enables analysis of the dynamics of each population. For the solar wind ion moments, we also avoid contamination by alpha particles by limiting the energy range of the particle distributions. These ions however, still appear in distribution spectrograms at energies twice the solar wind beam energy. We also note that the FPI energy and angular sampling is not optimized to resolve the cold and narrow solar wind beam. It is prone to under estimate the density and over estimate the temperature moments. We obtain the solar wind temperature and plasma $\beta$ from the OMNI dataset.

AC-Coupled magnetic field variations are measured by the search-coil magnetometer (SCM) instrument \citep{LeContel2016TheMMS}, while 3D AC-coupled electric field measurements are provided by double probe sensors in the spin plane \citep{Lindqvist2016TheMMS}, and on the spin axis \citep{Ergun2016TheMission}, collectively known as electric field double probe (EDP). The AC-Coupled electric and magnetic field fluctuations are scanned at 8192 samples per second. The EDP data have been filtered and reprocessed to remove the spacecraft spin effects. Dynamic power spectral densities (PSDs) of electric and magnetic field fluctuations are generated by applying a Fourier transform on SCM and EDP time series data with a 31.2 ms sliding window.


\subsection{Timing Method}
The MMS mission consists of four spacecraft. Using measurements at four corners of the MMS tetrahedron, the speed and propagation direction of a plasma structure are calculated from \citep{Russell1983MultipleNormals, schwartz_shock_1998}: 
\begin{equation}
   \begin{bmatrix}
       \delta r_{12} \\
       \delta r_{13} \\
       \delta r_{14}
   \end{bmatrix} 
   \frac{\hat{\textbf{k}}_s}{|\textbf{V}_s|} = 
   \begin{bmatrix}
       \delta t_{12} \\
       \delta t_{13} \\
       \delta t_{14}
   \end{bmatrix} 
\end{equation}

\noindent where $\delta r_{ij}$ are the inter spacecraft distances, $\hat{\textbf{k}}_s$ and $|\textbf{V}_s|$ are the propagation unit vector and the speed of the structure, and $\delta t_{ij}$ are the time lags determined by cross correlating the measurements between spacecraft pairs. This method is based on the assumption of a planar structure, meaning that there can be uncertainties in the results of this method if the structure substantially changes on timescales shorter than the transition time between spacecraft.

\subsection{Shock Normal Calculations}
To calculate the shock normal vector at the point of the crossing we used a conic section model of \citet{Peredo1995Three-dimensionalOrientation} fit to the crossing location of spacecraft. From this method, we obtain $\hat{\textbf{n}} = (0.96, -0.27, -0.05)$ in the GSE coordinates. We can also estimate $\hat{\textbf{n}}$ from the mixed mode coplanarity method \citep{schwartz_shock_1998}:

\begin{equation}
\hat{\textbf{n}} = \pm \frac{(\Delta \textbf{B} \times \Delta \textbf{V}) \times \Delta \textbf{B}}{|(\Delta \textbf{B} \times \Delta \textbf{V}) \times \Delta \textbf{B}|}
\end{equation}

\noindent where $\Delta \textbf{B}= \textbf{B}_\text{down} - \textbf{B}_\text{up}$ is the difference between downstream ($\textbf{B}_\text{down}$) and upstream ($\textbf{B}_\text{up}$) magnetic field, and similarly $\Delta \textbf{V}$ is the difference in the plasma bulk flow velocity. The upstream magnetic field and flow velocity in the pristine solar wind are listed in Table~\ref{table:1} (i.e., $\textbf{B}_\text{up} = \textbf{B}_{\text{IMF}}, \textbf{V}_\text{up} = \textbf{V}_{\text{SW}}$). For the downstream values, the magnetic field and the ion velocity moment data are averaged between 03:58:52 and 03:59:00 UT, to give $\textbf{B}_\text{down} = (-3.21, -13.15, 5.1)$ nT, $\textbf{V}_\text{down} = (-132, -65, -48)$ kms\textsuperscript{-1}. From Equation (2) we obtain $\hat{\textbf{n}} = (0.96, -0.24, -0.10)$. The vectors are in GSE coordinates unless marked otherwise. The two estimates of $\hat{\textbf{n}}$ are in agreement within $3^\circ$, indicating that the shock normal estimate is reasonably accurate, though we use the normal vector obtained from the model in our analysis. We define the orthogonal shock normal coordinate basis (NCB) with $\hat{\textbf{n}}$, $\hat{\textbf{t}}_1$, and $\hat{\textbf{t}}_2$ vectors. These basis vectors are related such that $\hat{\textbf{t}}_2 = \hat{\textbf{n}} \times \hat{\textbf{B}}_{up}$, and $\hat{\textbf{t}}_1$ is perpendicular to $\hat{\textbf{n}}$ and $\hat{\textbf{t}}_2$. $\hat{\textbf{t}}_1$ and $\hat{\textbf{t}}_2$ are tangent vectors to the shock surface, $\hat{\textbf{n}}$ and $\hat{\textbf{t}}_1$ form the coplanarity plane, while $\hat{\textbf{t}}_2$ is along the upstream motional electric field. We use the normal incidence frame (NIF) to study the ion dynamics. In the NIF frame, the solar wind flow is along the shock normal vector, and the the frame's velocity is obtained from: $\textbf{V}_{\text{NIF}} = \hat{\textbf{n}} \times (\textbf{V}_{\text{SW-sh}} \times \hat{\textbf{n}})$, where $\textbf{V}_{\text{SW-sh}}$ is the pristine solar wind velocity in the shock rest frame. For a nonstationarity shock front dynamically modulated by reformation cycles, the timing method cannot be applied to measure the shock parameters. The shock speed along the normal vector ($V_{shock,n}$) can be calculated based on the traverse time over the shock foot by a specularly reflected solar wind ion \citep{gosling_specularly_1985}. In the absence of more accurate methods, this simple approach provides a good rough estimate of the shock speed.

\begin{deluxetable*}{p{6cm} c}
\tablecaption{Upstream plasma and shock parameters. \label{table:1}} 
\tablehead{
\colhead{Parameter} & \colhead{Value}
}
\startdata
Flow velocity $\textbf{V}_\text{SW}$  & (-464, 31, -14) kms\textsuperscript{-1} \\ 
Magnetic field $\textbf{B}_\text{IMF}$  & (-0.80,-1.97,1.55) nT \\ 
Plasma density $n_{SW}$ & 10.5 cm\textsuperscript{-3}\\
Ion temperature $T^*_{i,SW}$ & 2.33 eV \\
Plasma $\beta$\textsuperscript{*}  & 9.1 \\
Ion cyclotron frequency $f_{ic}$ & 0.048 Hz  \\
Ion plasma frequency $\omega_{pi}$ & 664.6 Hz \\
Ion inertial length $\lambda_i = c/\omega_{pi}$ & 72 km \\
Electron cyclotron frequency $f_{ec}$ & 74.3 Hz  \\
Alfv\'en Mach number $M_{A}$ & 27 \\
Fast magnetosonic Mach number $M_{fm}$ & 15  \\
Shock normal vector $\hat{\textbf{n}}$ & (0.96, -0.27, -0.05) \\
Shock angle $\theta_{Bn}$  & $83^{\circ}$ \\
$V_{shock,n}$ & 21.7 kms\textsuperscript{-1} \\
\enddata
\tablecomments{\textsuperscript{*}From OMNI dataset. Vector quantities are in the Geocentric Solar Ecliptic (GSE) coordinates, in which $+x$ is toward the Sun and $+z$ is normal to the ecliptic plane.}
\end{deluxetable*}

\section{Observations}
\label{sec:obs}

\begin{figure}[htb]
\centering
\plotone{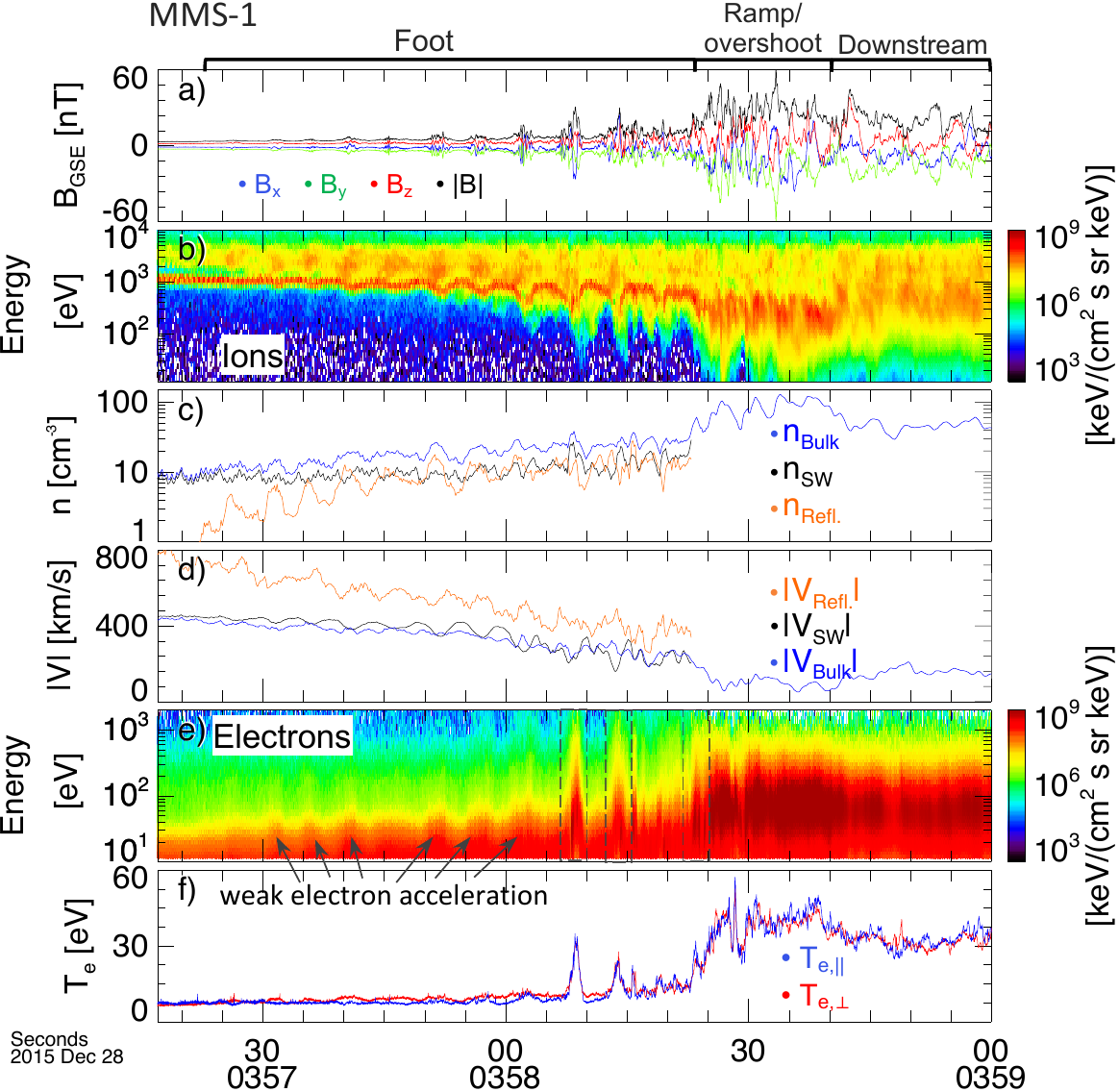}
\caption{Overview of the shock crossing event on 28 December 2015: (a) magnetic field magnitude; (b) ion energy flux spectrogram in logarithmic scale; (c) density of solar wind ions (black), reflected ions (yellow), and bulk plasma (blue); (d) ion velocity moment of solar wind beam, reflected ions, and bulk plasma flow; (e) electron energy flux spectrogram; amd (f) electron temperature parallel (blue) and perpendicular (red) to the local magnetic field. Data in panels (a - f) are from MMS-1 spacecraft. $R_E$: Earth radius.}
\label{Fig:Overview}
\end{figure}

The bow shock crossing event is observed by MMS on 28 December 2015 between 03:56:12 and 04:00:15 UT at (10.9, -4.9, -1.1) R\textsubscript{E}. The upstream IMF is relatively weak but steady during this time. The solar wind beam temperature is cold, as measured by the width of the ion beam, while the solar wind density is high. The shock Alfv\'enic ($M_A$) and fast magnetosonic ($M_{fm}$) Mach numbers are recorded at 27 and 15, respectively, and the shock angle is $\theta_{Bn} = 83^\circ$. Table \ref{table:1} includes a list of upstream plasma and shock parameters. Figure~\ref{Fig:Overview} shows an overview of plasma and field data across the shock measured by the MMS spacecraft 1 (MMS-1) from 03:57:15 UT in the solar wind, until 03:59:00 UT when the spacecraft is in the magnetosheath. Panel (a) shows the magnetic field component and strength. Different regions of the shock are annotated on this panel. We will include the magnetic field profile in other time series figures for context. The ion energy spectra in panel (b) shows that solar wind protons form a narrow beam with a bulk energy $\sim 1150$ eV. We observe populations throughout the foot of accelerated ions above the solar wind beam energy, which are identified as solar wind ions reflected by the shock. The maximum energy of these ions is capped at $\sim 5$ keV, indicating that the efficiency of the acceleration mechanism (or mechanisms) driving these these ions is relatively constant during this period.

Separate density and velocity moments for the solar wind (black), reflected ions (yellow), and the bulk plasma (blue) are shown in panels (c) and (d). Densities are plotted on a logarithmic scale axis to highlight differences between ion populations. The shock ramp and overshoot regions are characterized by a significant heating and deceleration of the solar wind beam, distinct increase in the plasma density, and increased electron temperature that continues to the downstream. Once inside the ramp, the solar wind plasma experiences significant heating and the core solar wind population is no longer observed. Upstream of the shock, periodic bulk electron acceleration are observed as annotated in panel (e). Closer to the shock front, major bulk electron heating, comparable to magnetosheath levels, occurs at 03:58:09 UT, followed by two closely distanced events at around 03:58:15 UT. During these instances, boxed with dashed rectangles, we observe significant isotropic electron energization, evident by the increase in both parallel (blue) and perpendicular (red) temperature components in panel (f).

The foot region of supercritical quasi-perpendicular shocks typically exhibits a gradual increase in the magnetic field strength caused by reflected ions. In the foot region in Figure~\ref{Fig:Overview}, we also observe quasi-periodic and isolated enhancements in the magnetic field strength ($|\textbf{B}|$) which, particularly closer to the shock, are as strong as the downstream magnetic field strength. The enhancements are correlated with increases in the density of reflected ions (the yellow line in panel (c)), and with momentary slowdown in the solar wind flow. These quasi-periodic modulations are consistent with (partial) reformation cycles of the shock \citep{sulaiman_quasiperpendicular_2015,sundberg_dynamics_2017, Madanian2020NonstationaryMars}.

\begin{figure}[ht]
\centering
\includegraphics[width=0.5\linewidth]{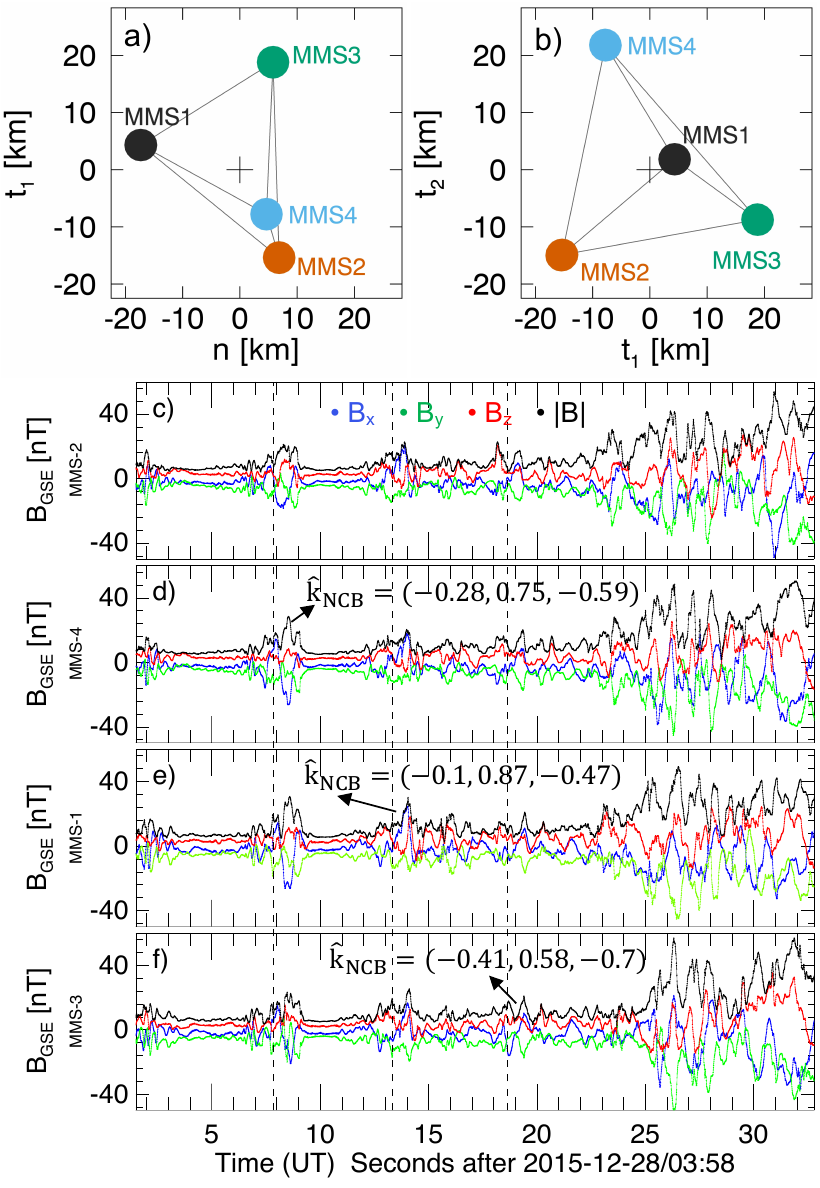}
\caption{Panels (a) and (b) show the MMS tetrahedron 
formation in $n - t_1$ and $t_1 - t_2$ planes of NCB coordinates, respectively. Panels (c - f) show the magnetic field data from spacecraft 2, 4, 1, and 3, respectively. Dashed vertical lines are drawn on three reformation cycles when a major sign change in B\textsubscript{x} is observed in MMS-2 data. $\hat{\textbf{k}}_\text{NCB}$ show the propagation direction of each wave packet in NCB coordinates. Structures are in propagation towards the shock and along the magnetic field.}
\label{Fig:Overview_nonst}
\end{figure}

Top two panels in Figure~\ref{Fig:Overview_nonst} show four MMS spacecraft tetrahedron in NCB coordinates. Magnetic field data from all four MMS spacecraft for a sub-interval focused on the shock front are shown in panels (c - f). Dashed vertical lines are drawn on three reformation cycles upstream of the shock. A specific characteristics of the wave packets (i.e., sign change in the $B_x$ component) is first seen in MMS-2. Moments later, a similar signature appears in MMS-4, then MMS-1, and finally in MMS-3 data. Using the timing method, we determine the propagation direction of the feature at around 03:58:08 UT $\hat{\textbf{k}_{NCB}}=(-0.28, 0.75, -0.59)$ in GSE coordinates, consistent with the order of observations. The structure is in propagation at an angle of $\sim 28^{\circ}$ with the background magnetic field. The propagation direction of other cycles are also annotated on the figure. The main shock transition layer (i.e., ramp and overshoot) in this figure begins at $\sim$ 03:58:24 UT and is characterized by large amplitude fluctuations in all components of the magnetic field. A structure similar to the previous cyclic enhancements is observed between 03:58:22-24 UT in MMS-2, MMS-4, and MMS-1 ,consistent with the order of observation of previous cycles, but it is missing in MMS-3 data. If these were shock front breathing motion or upstream propagating whistler waves, given the close alignment of MMS-2, MMS-3, and MMS-4 along $\hat{\textbf{n}}$, one would expect to see changes in magnetic field data in all three spacecraft at about the same time. However, data in Figure \ref{Fig:Overview_nonst} indicate that the enhancements are generated upstream and their propagation vector has an Earth-ward component, and they significantly modulate the shock front.

\subsection{Shock Reformation}

\begin{figure}[htb]
\centering
\plotone{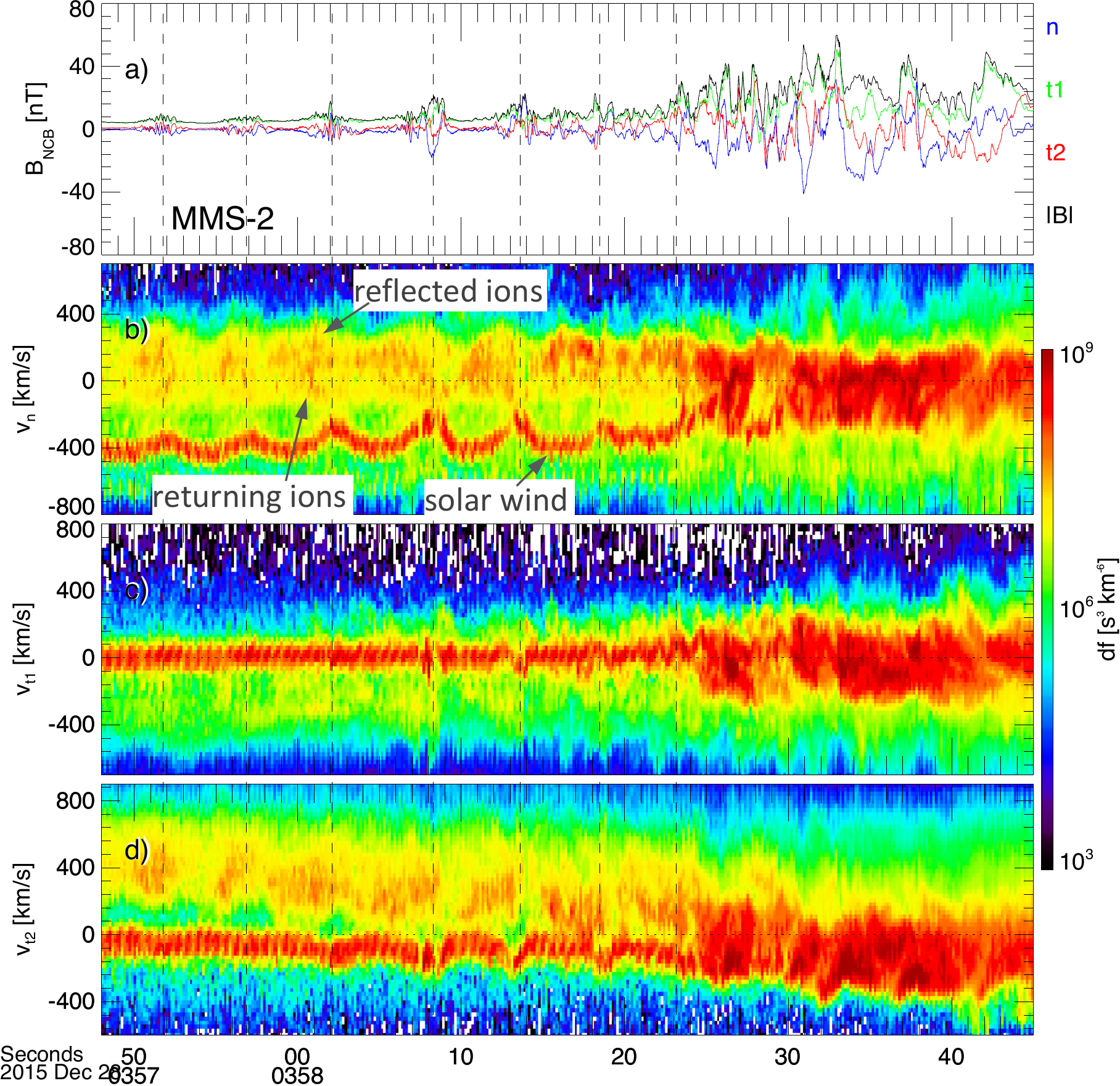}
\caption{The MMS-2 spacecraft observations revealing the fine scale ion dynamics across the shock. Panel (a) shows the magnetic field in the NCB (Section \ref{sec:meth}). Panels (b - d) show ion velocity distributions in the NIF and along $\hat{\textbf{n}}$, $\hat{\textbf{t}}_1$, and $\hat{\textbf{t}}_2$, respectively. Vertical dashed lines mark the reformation cycles in the foot. Different ion populations are annotated in panel (b).}
\label{Fig:Reflec}
\end{figure}

Reformation cycles observed in the foot have a period of $\sim 5.5 \text{ s } \sim 0.22f_{ic}^{-1}$, consistent with previous observations \citet{sundberg_dynamics_2017} and \citet{sulaiman_quasiperpendicular_2015}. They appear to be a proton/subproton scale effect. Figure~\ref{Fig:Reflec} shows ion distributions in the velocity space near the foot region leading to the shock ramp as observed by MMS-2. In panel (a) magnetic field vectors rotated to NCB coordinates are shown. Quasi-periodic reformation sequences are identified with vertical dashed lines, corresponding to the end points of half-filled vortices in the $V_n$ distribution spectrogram. The ion data in panels (b - d) are in the NIF moving with a velocity of \textbf{V}\textsubscript{NIF} = (-34.89, -112.35, -55.8) kms\textsuperscript{-1} in the shock rest frame. The solar wind beam at $V_n \sim -400$ kms\textsuperscript{-1} is evident in panel (b). Reflected ions with positive $V_n$ velocities are observed, as well as another population with a small negative $V_n$ component. The latter is identified as reflected ions that have already been turned around by the motional electric field. Panel (c) shows ion distributions along $\hat{\textbf{t}}_1$. Since both solar wind and reflected ions travel at nearly perpendicular angles to \textbf{B}, their projected velocity along \textbf{B} is zero. The last panel shows the $V_{t2}$ distribution of ions. Farther upstream from the shock, reflected ions have higher $V_{t2}$ velocities as they have spent a longer time in the upstream motional electric field. The trace of reflected ion in the velocity space (not shown) farther from the shock more closely falls along the predicted trajectory of a specularly reflected ion \citep{Paschmann1982, Madanian2020NonstationaryMars}.

For reformation cycles at the beginning of the interval near the upstream edge of the foot, pileup of reflected ions are collocated with the magnetic field enhancements. Closer to the shock, solar wind and reflected ions have been interrupted by the reformation cycles, as seen in distributions near the last four vertical lines in panel (d). These last cycles also show markedly higher bulk electron acceleration and we observe solar wind beam compression for the first time at 03:58:08 UT. These observations indicate that buildup of reflected ions at the upstream edge of the foot does not instantly generate a new shock front but they are observed quasi-periodically in the foot. Closer to the shock, they begin to exhibit shock-like plasma heating and acceleration, and interrupt both the solar wind and reflected ions from the main shock. We refer to these cycles as ''proto-shocks'', as they do exhibit shock-like behavior, but are not strong or expanded enough to fully replace the main shock.

We also note that in Figure~\ref{Fig:Reflec}.b after the last reformation cycle at 03:58:23 UT, unlike previous cycles, the solar wind beam continues to decelerate to $V_n \sim 0$. At this time, we observe strong ion reflection with intensities higher than previously observed, while ion distributions in panels (c) and (d) show a heated solar wind plasma. These features are significantly different than those of the upstream reformation cycles, and the measurements at 03:58:25 UT most likely represent the first encounter with the main shock layer. In other words, while reformation occurs in the foot, the main shock layer (or the boundary at which the main shock properties are observed) never disappears. The solar wind flow gradually decreases throughout the foot. But as evident in Figure~\ref{Fig:Overview}.d, solar wind bulk flow speeds are consistently higher than the downstream bulk plasma flow speed. In addition, reflected ions are constantly present in the foot except when interrupted by the proto-shocks. These ions also seem to occupy a broad range of positive $V_n$ velocities, rather than being a beam of ions. This can be due to simultaneous reflection taking place from two reflection surfaces, non-specular reflection \citep{Sckopke1983}, or interaction with upstream proto-shocks. The shock surface may also be rippled, leading to reflection at different directions and causing reflected ions to be at different stages of their gyration and consequently different energy levels by the time they arrive at the location of the MMS spacecraft \citep{Ofman2013RippledNormals}. 2D particle-in-cell (PIC) simulations have shown that both shock reformation and shock surface rippling can affect the excitation of the electrostatic waves in the foot by modifying the intensity of reflected ions at different locations across the shock \citep{Hao2016IONSIMULATIONS, Matsukiyo2003ModifiedShocks}.

\subsection{Surface Ripples}

\begin{figure}[htb]
\centering
\plotone{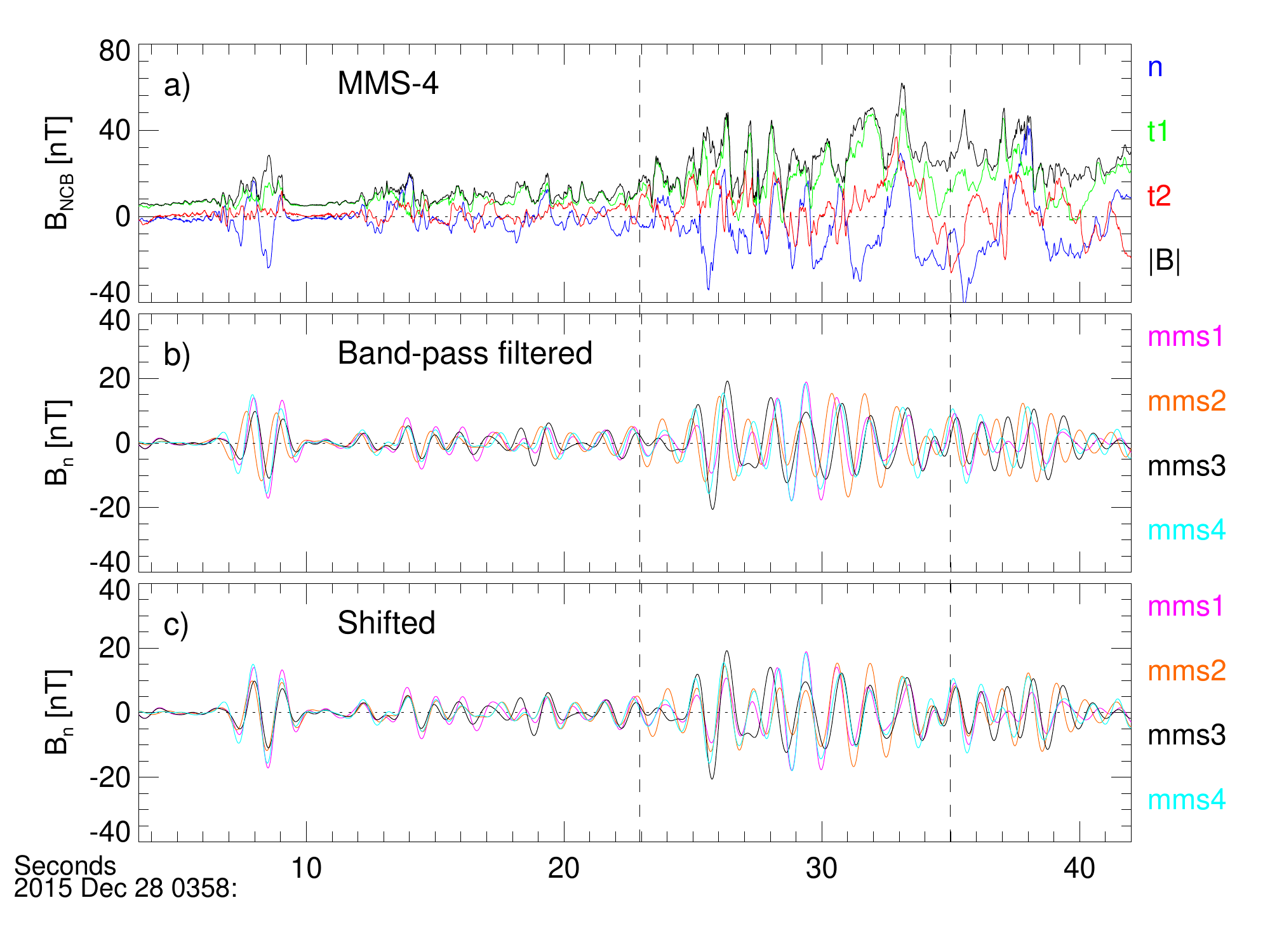}
\caption{Analysis of shock ripple properties via the timing method: a) Magnetic field magnitude and components in the NCB from spacecraft 4; b) normal component of the magnetic field $B_n$ from all four spacecraft, data are band-pass filtered in the (0.56-1.22) Hz frequency range; and c) $B_n$ signals from spacecraft 2, 3, and 4 are shifted in time to account for time delays in the observations with respect to spacecraft 1. The vertical dashed lines specify the time period used for cross correlation and identification of time lags.}
\label{Fig:Ripples}
\end{figure}

During this event, sharp magnetic gradient of the ramp, typically present at quasi-perpendicular shocks, is replaced by large amplitude magnetic oscillations. 
As observed in MMS-4 data shown in Figure~\ref{Fig:Ripples}.a, the normal component of the magnetic field ($B_n$) periodically reverses sign during this period, which can be attributed to a rippled shock surface \citep{lowe_properties_2003, johlander_rippled_2016}. $B_n$ data band-pass filtered in the 0.56 - 1.22 Hz frequency range from all four spacecraft are shown in panel (b). The selected frequency range covers most of the high amplitude variations we observe, but does not include the lower frequency variations due to upstream reformation cycles ($\sim 0.2$ Hz). By cross correlating the signals, we obtained time lags $\delta t_{12} = -210.9$ ms, $\delta t_{13} = 23.4$ ms, and $\delta t_{14} = -23.4$ ms between measurements in spacecraft 1 and the three other spacecraft. Using the timing method described in Section~\ref{sec:meth}, we find the phase speed of the wave in the spacecraft frame $V_{ph-sc}=136$ kms\textsuperscript{-1}. The wave propagates mostly along the shock surface with the wave vector $\hat{\textbf{k}_{NCB}}$ = (-0.22, 0.80, -0.54). After correcting for the Doppler effect using the locally measured plasma velocity, the wave phase speed in the local plasma rest frame is $\sim 41$ kms\textsuperscript{-1} or $\sim 0.6 V_{A}$, where $V_{A}$ is the average local Alfv\'en speed. The wave's characteristic wavelength is $\lambda_{wave} = V_{ph-sc}/f_{sc} = 153$ km or $2.1 \lambda_i$, where $\lambda_i$ is the upstream ion inertial length. These properties are consistent with surface ripples and large amplitude ion-scale waves generated by AIC instability \citep{lowe_properties_2003, Davidson1975ElectromagneticPlasmas}. Waves have an elliptical polarization, and are far below the lower hybrid frequency. There exist some deviations in the shifted signals, particularly for spacecraft 3, indicating that other processes with similar frequencies could be in play. Nonetheless, the good alignment of the shifted signals verifies that the time lags are properly determined.  

Similar waves with comparable amplitude are present around the reformation cycle at the beginning of the interval between 03:58:05 and 03:58:05 UT. It seems that the large amplitude waves creating the surface ripples are also present within the incoming proto-shock, which suggests that rippling begins to develop in upstream proto-shocks. Farther upstream, the wave amplitude becomes very small. It is worth noting that the period of ripple waves is quite different than the periodicity of the reformation process, which enables to distinguish the two effects. Numerical simulations by \citet{gingell_mms_2017} has shown a similar scenario, though with different shock parameters. The authors showed that surface ripples are modulated by the periodic reformation of the shock front, and transient ripples develop at the newly formed shock on timescales shorter than the upstream ion gyroperiod. This further signifies our observations and the feasibility of our interpretations. Similar processes have also been observed for oblique shocks \cite{lefebvre_reformation_2009}, and have been investigated for astrophysical quasi-parallel shocks through numerical simulations \citep{Caprioli2014SIMULATIONSAMPLIFICATION, Caprioli2015SimulationsShocks}. Our observations of this process at a nearly perpendicular shock suggest that such shock reformation/generation process is a global phenomenon that can occur for a wide range of shock parameters.

\subsection{Upstream Electromagnetic Perturbations}

In this section we examine electrostatic and electromagnetic wave activities upstream of, and within the shock layer. We restrict our discussion to identifying wave features associated with reformation cycles, noting that each structure has internal wave characteristics quite different than the others. Characterizing the nature and the generation mechanism of all these waves, although critical, is beyond the scope of this paper.

\begin{figure}[htb]
\centering
\plotone{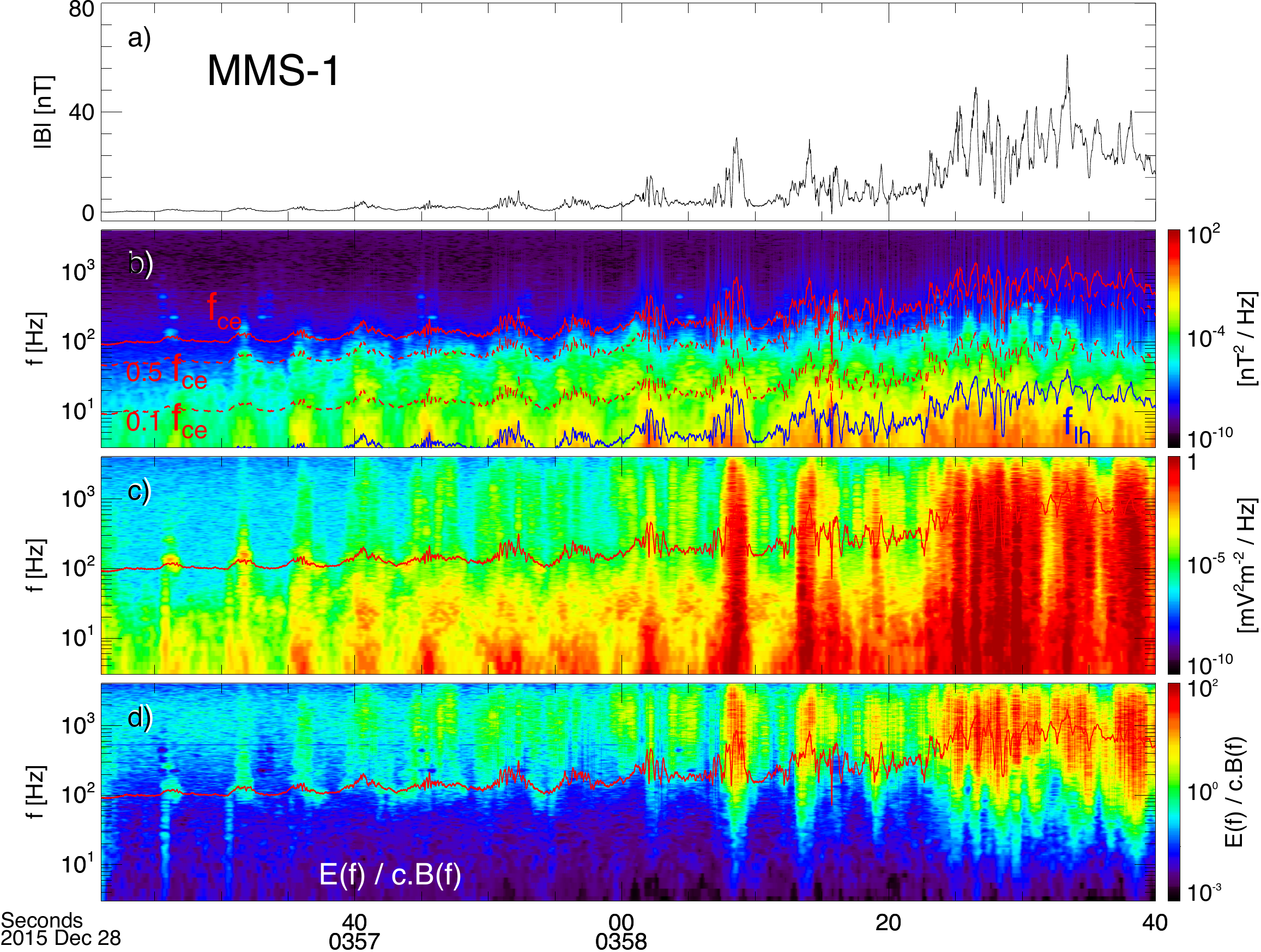}
\caption{Electromagnetic wave activity in the foot region observed by MMS-1: a) Magnetic field profile; b) power spectral density (PSD) of magnetic field fluctuations; c) PSD of electric field fluctuations; and d) ratio of electric to magnetic field fluctuations in the frequency domain ($E(f) / cB(f)$, the conversion factor $c$ is the speed of light). The red solid line on panel (b) shows the electron cyclotron frequency $f_{ce}$. $0.5 f_{ce}$ and $0.1 f_{ce}$ frequency lines are shown with red dotted lines. The $f_{ce}$ line is repeated on panels (c) and (d) for reference. The blue line on panel (b) is the lower hybrid frequency $f_{lh}$.}
\label{Fig:WavePSD}
\end{figure}

Figure~\ref{Fig:WavePSD} shows the dynamic power spectral densities (PSDs) of electric and magnetic field perturbations. In panel (a) the magnetic field profile is shown as a reference. Panels (b) and (c) show, respectively, the electric and magnetic field PSDs in the $3 - 4096$ Hz frequency range. Bursts of electric and magnetic fluctuations at frequencies around and below the lower hybrid frequency ($f_{lh}$, the blue line in panel (b)) are observed. These perturbations upstream of the ramp are correlated with the reflected ion densities and solar wind decelerations discussed in Figure~\ref{Fig:Overview}. The total electric and magnetic wave powers (integrated over the whole frequency range) oscillates with the same period as the shock reformation. The shock transition layer, between 03:58:24 UT to the end of the interval, is distinguished by high power broadband electric perturbations.
The ratios of the electric to magnetic field fluctuations are shown in panel (d). Yellow and red colors correspond to ratios much greater than 1, indicative of electrostatic waves. As expected, strong electrostatic waves are present within the main shock transition layer \citep{Scudder1995AShocks, Bale1998BipolarHoles, bale_quasi-perpendicular_2005, Vasko2018SolitaryShocks, goodrich_mms_2018, wang_electrostatic_2020, wilson_quantified_2014-1}. We also see similar electrostatic waves upstream of the shock layer inside the last three reformation cycles, which distinguishes them from the earlier sequences. These cycles, or proto-shocks, also show significant bulk electron heating (Figure~\ref{Fig:Overview}.f). This is an important observation as it reveals that not all cyclic enhancements exhibit shock-like properties, and bursty high frequency electrostatic waves are not restricted to the main shock layer. 

Our wave analysis also shows that the sporadic high frequency magnetosonic waves between $0.1 f_{ce}$ and $0.5 f_{ce}$ (shown with red dotted lines in panel (b)) are circularly right-hand polarized and in propagation quasi-parallel to the background magnetic field (wave angle $<30^{\circ}$). The waves are consistent with a source of electron temperature anisotropy ($T_{e,\perp}/T_{e,\parallel}>1$, see Figure \ref{Fig:Overview}.f) \citep{Kennel1966LimitFluxes,Gary1996WhistlerBound}, which is likely created by magnetic increases during the reformation process, and associated gyrobetatron effects.

At the beginning of the interval in Figure~\ref{Fig:WavePSD}.c, isolated high frequency quasi-electrostatic fluctuations near the electron cyclotron frequency ($\sim100$ Hz) are observed. Their generation mechanism could be related to reflected ions. They are generated near the upstream edge of the foot, where specularly reflected ions are accelerated along the motional electric field and travel almost perpendicular to the magnetic field. The ion gyration trajectory at that point can be considered as almost a straight line, hence providing a nonmagnetized fast ion component that destabilizes the plasma and causes generation of high frequency quasi-electrostatic waves \citep{Muschietti2013MicroturbulenceShocks,sundberg_dynamics_2017, Omidi1987AShocks}. More investigations are required to better identify the nature of the instability. 

\begin{figure}[htb]
\centering
\includegraphics[width=0.7\linewidth]{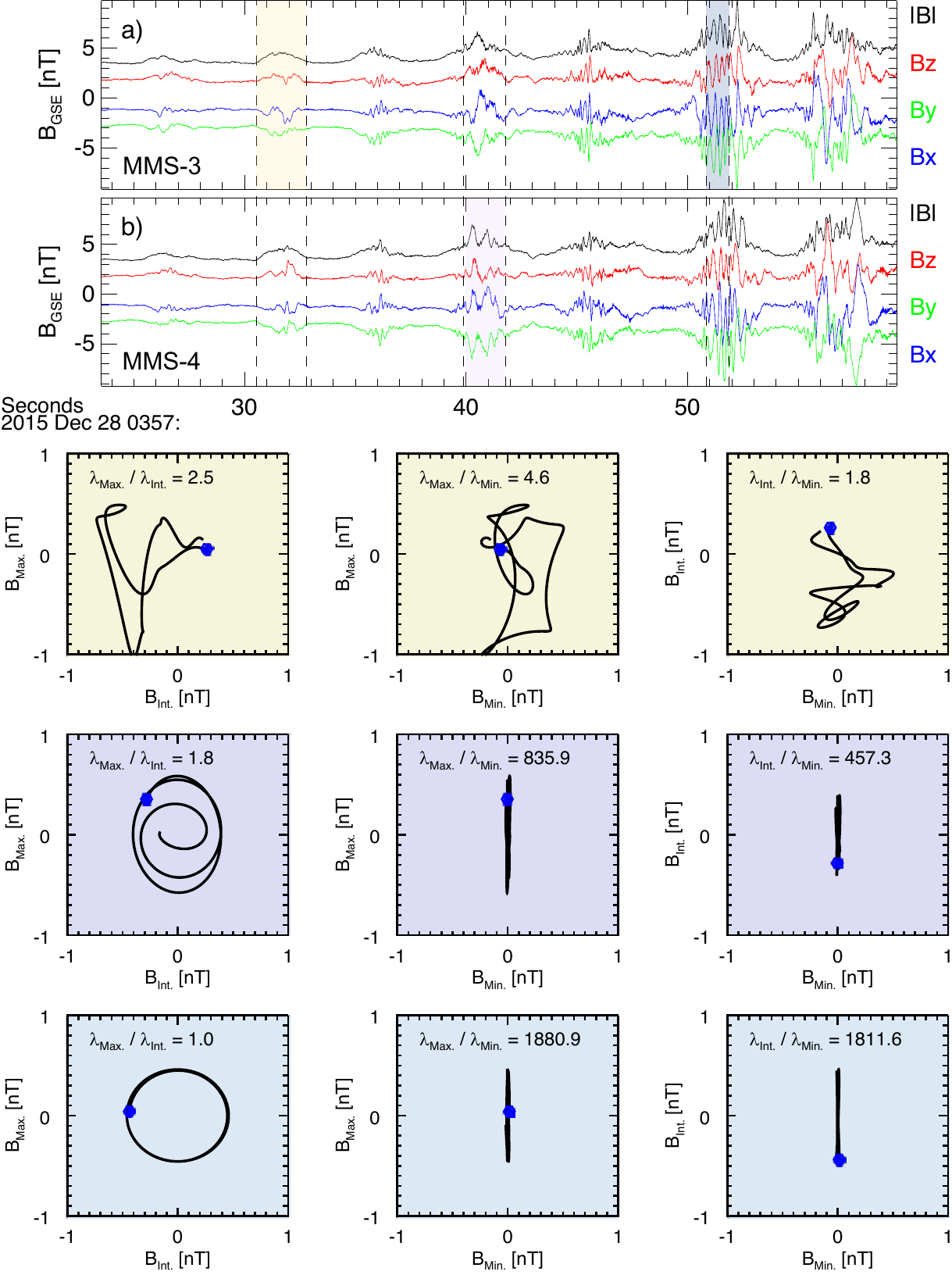}
\caption{Properties of magnetic fluctuations. Panels (a) and (b) show the magnetic field components and magnitude measured by MMS-3 and MMS-4, respectively. The first row of hodograms at the bottom correspond to the first time interval labeled with the yellow box for MMS-3. The second row corresponds to the second purple interval for MMS-4, and third row shows the principal components of the magnetic field for the blue interval on MMS-3. Hodograms show the background subtracted, low pass filtered ($< 20$ Hz) data. The background magnetic field is pointing into the $\text{B}_{\text{Max.}} - \text{B}_{\text{Int.}}$ plane (the first column), and the blue dots mark the beginning of each interval. The band-pass frequency range is (0.1 - 4) Hz for the first interval, (1.4 - 1.8) Hz for the second interval, and (1.9 -2.1) Hz for the third interval.}
\label{Fig:nonlinear}
\end{figure}

Propagation and evolution of the upstream cyclic enhancements throughout the foot is rather nonlinear. Each enhancement is accompanied by a burst of low frequency waves, some of which have characteristics consistent with the whistler wave mode, while some show no particular polarization. To verify this nonlinear pattern, we perform wave analysis on magnetic field data from all spacecraft, over various frequency ranges and time periods. Three illustrative examples are discussed in Figure~\ref{Fig:nonlinear}, though many other frequency bands were examined. Panels (a) and (b) show magnetic field data from MMS-3 and MMS-4 between 03:57:24 and 03:57:59 UT. We apply the minimum variance analysis on select intervals highlighted with yellow and blue in MMS-3 and purple in MMS-4 time series. Hodograms at the bottom show band-pass filtered, background subtracted field variances for $\text{B}_{\text{Max.}} - \text{B}_{\text{Int.}}$ in the first column, $\text{B}_{\text{Max.}} - \text{B}_{\text{Min.}}$ in the second column, and for $\text{B}_{\text{Int.}} - \text{B}_{\text{Min.}}$ in the third column. The ratios of corresponding eigenvalues are annotated on each panel, and the background magnetic field points into the page of the $\text{B}_{\text{Max.}} - \text{B}_{\text{Int.}}$ planes. During the first interval (yellow segment) we find no identifiable wave pattern at any frequency range in MMS-3 data. This is evident in the first row hodograms which show field variations in the 0.1 - 4 Hz frequency range in that period. For the second interval (purple segment), right-handed elliptically polarized waves in the 1.4 - 1.8 Hz frequency range are observed. In the last interval, using MMS-3 data, we observe 2-Hz waves with right-handed circular polarization, the typical signatures of the whistler mode waves. By comparing the wave activity in the middle interval (03:57:40 - 03:57:42 UT) in MMS-3 and MMS-4, we notice a small frequency shift in the high amplitude waves. In addition, the last wave packet around 03:57:57 UT, the amplitude of the magnetic peak in MMS-3 data shows a decrease compared to that in MMS-4, while the position of the peak has also changed within the cycle. Note the separation between MMS-3 and MMS-4 along the magnetic field is about 25 km ($\sim0.3\lambda_i$).

Overall, we find that close to the shock, waves observed by all spacecraft share the same wave normal angle distribution (irrespective of the $180^\circ$ ambiguity). Far from the shock, each spacecraft sees distinct wave characteristics and the wave distributions appear to switch between relatively high and relatively low wave normal angles, and this behavior intensifies for 2-10 Hz waves. Not only polarization, amplitude, and duration of waves change from one cycle to another, waves also substantially evolve during the short travel between spacecraft. The nature of instabilities also varies from one cycle to another, showing complex and nonlinear evolution of wave packets as they propagate in the foot. These variations however, all begin with modest magnetic enhancement in the IMF generated by reflected ions. They transform into proto-shocks as they propagate Earth-ward. 

These signatures are inconsistent with ultra-low frequency (ULF) waves which have circular polarization and a period similar to the upstream ion gyroperiod. The waves are also inconsistent with ion Weibel instability (IWI) which generates linearly polarized waves. Interaction of reflected ions with incoming solar wind electrons or ions can cause foot instabilities that excite waves in the whistler mode branch. Modified Two Stream Instability (MTSI) due to relative drift between reflected ions and incoming solar wind electrons (fast drift), and incoming solar wind ions and electrons (slow drift) has been frequently considered \citep{marcowith_microphysics_2016, muschietti_two-stream_2017, Matsukiyo2003ModifiedShocks, Umeda2012ModifiedSimulations, comisel_non-stationarity_2011, WilsonIII2016LowShocks, Hull2020MMSTransport}. This instability however, if excited, creates significant ion heating throughout the foot and suppresses the reformation process \citep{Shimada2005EffectBehavior,Matsukiyo2006OnShocks}, rather than creating episodic enhancements that we show in the foot. Furthermore, \citet{Gary1987Ion/ionFrequency} indicated that (fast drift) MTSI becomes dominant at low electron beta ($\beta_{e} < 0.5$), while at higher $\beta_e$ more resonant electrons stabilize this instability through increased electron Landau damping. Electron data for the time period we discussed in this paper show $\beta_e \geq 1.2$, and therefore fast drift mode MTSI is most likely not significant. The slow drift mode of MTSI could be a more viable candidate at high $\beta$ plasmas. Wave properties around 1.6 Hz in the middle interval (purple segment) of Figure~\ref{Fig:nonlinear}, indicate that the wave is in propagation towards the ramp ($\hat{\textbf{k}}_{\text{GSE}}= -0.66, -0.71, 0.22$) with $V_{ph-sc}=34$ kms\textsuperscript{-1} and $\lambda_{wave}= 21.4$ km $\sim 12 \lambda_e$, where $\lambda_e$ is the upstream electron inertial length. The plasma rest frame frequency of the wave is about 8 Hz $\sim 2.5 f_{lh}$. Since these characteristics are somewhat consistent with model predictions for drift mode of MTSI \citep{muschietti_two-stream_2017}, we do not rule out the possibility of some waves at certain frequencies and during some intervals being generated by the slow drift mode of MTSI.

\section{Conclusions and Discussion}
By studying the dynamics of a quasi-perpendicular shock at very high Mach numbers, we are able to isolate and investigate the effects of ion dynamics on the shock structure. The high upstream Mach number of the upstream flow does not allow for shock-generated dispersive whistler waves to propagate throughout the foot and the shock dynamics are tied to the reflected ion dynamics. 

We observe signatures of shock reformation in the form of magnetic enhancements that evolve in a nonlinear manner to form proto-shocks as they traverse the foot. This is a different mechanism than previously suggested by simulations. What we observe here, instead, is that quasi-periodic reformation is indeed initiated by reflected ions through generation of modest magnetic enhancements at the upstream edge of the foot; however, the enhancement do not immediately replace the main shock. As they convect Earth-ward, they transform into proto-shocks through nonlinear amplification of electric and magnetic fields within the enhancements. This amplification is separate from the compression and amplification that occur at the main shock layer. The proto-shocks possess high frequency electrostatic waves and exhibit significant electron heating, and interfere with both the solar wind flow and reflected ions from the main shock.

We show that Alfv\'en Ion Cyclotron (AIC) waves within the shock ramp and overshoot form surface ripples with a wavelength of 2.1 ion inertial lengths, and an average period of $\sim 1.2$ s, which is the same as the local ion gyroperiod. They propagate along the shock surface with a phase speed of 0.6 times the local Alfv\'en speed. The rippling effect is also observed within some of the reformation cycles upstream of the main shock. 

The electric and magnetic field perturbations show the same periodicity as the reformation cycles. Even though shock-generated whistler waves are absent in the foot (nonlinear whistler critical Mach number $M_{nlw} \sim 3.7 \ll M_{fm}, M_{A}$), we observe intermittent whistler waves between $0.1 f_{ce}$ and $0.5 f_{ce}$ frequency range (Figure~\ref{Fig:WavePSD}.b). At the beginning of the interval, they coincide with reformation cycles, but later are observed in between cycles. These whistler waves are correlated with the electron temperature anisotropy. The most likely cause of the anisotropy is gyrobetatron effects associated with the increased magnetic fieldat reformation cycles. In Figure~\ref{Fig:nonlinear}, we show that some cycles also carry locally generated low frequency whistler waves.

Our observations are unique for high plasma $\beta$ shocks. This regime of shock parameters has been under studied by numerical simulations. We observe signatures of shock reformation with upstream plasma $\beta \sim 9$. Most simulation studies have indicated that at high $\beta$ ($>1$), shock reformation is suppressed \citep{hellinger_reformation_2002, hada_shock_2003, scholer_quasi-perpendicular_2003, Balogh2013PhysicsWaves}. Such thresholds in the models are due to limitation on the number of particles that can be included in each cell of the simulation box. Thus, high $\beta$ conditions are normally achieved by increasing the ion temperature, which causes a smooth but extended increase of the magnetic field in the shock foot. Unrealistic ion to electron mass ratios can also lead to overestimation of dispersive effects, which also work to increase the ion temperature \citep{lembege_nonstationarity_2009}. Imposing single population isotropic Maxwellian distribution for each particle species is rather unrealistic and definitely affects the instabilities present in the foot region of quasi-perpendicular shocks and foreshock region of quasi-parallel shocks. Nonetheless, models can provide interpretations from a different stand point, and a detailed simulation analysis for these observations is left for a future study. 

The high solar wind plasma $\beta$ in the event we discussed here is due to the high plasma density  and the very weak IMF strength, both of which also contribute to achieving the high Alfv\'en Mach number under the typical solar wind speed. Relatively weak upstream magnetic field is common in interstellar and astrophysical shocks \citep{Donnert2018MagneticSimulation, Petrukovich2019Low-frequencyShock}, and our observations and interpretations can, to some extent, be applied to those structures.

%
\acknowledgments  
We would like to thank D. Sibeck for helpful discussions. All data used in this study are publicly available via the MMS Science Data Center (\url{https://spdf.gsfc.nasa.gov/pub/data/mms}), and NASA/GSFC’s OMNIWeb service (\url{http://cdaweb.sci.gsfc.nasa.gov}) for solar wind data. Data access and processing software includes the publicly available SPEDAS package \citep{Angelopoulos2019TheSPEDAS}. This work was supported in part by the NASA Award Number 80NSSC18K1366. L.B.W. acknowledges partial support through an International Space Science Institute (ISSI) team. A.L.B. is supported by NASA grant 80NSSC20M0189. The French LPP involvement for the SCM instrument is supported by CNES and CNRS.

\bibliography{Mendeleyreferences.bib}

\end{document}